\documentstyle[aaspp4,11pt]{article}

\begin{document}

\title{Why Does NGC 1068 Have a More Powerful Active Galactic Nucleus
than NGC 4258 ?}

\author{Takashi {\sc Murayama}  and Yoshiaki {\sc Taniguchi}\\
{\it Astronomical Institute, Tohoku University, Aoba,
Sendai 980-77} \\
{\it E-mail (TM): murayama@astr.tohoku.ac.jp}}

\author{(Received 10 April 1997;\hspace*{1em}accepted 9 May 1997)}
\author{To appear in Publications of the Astronomical Society of Japan}

\begin{abstract}
The nuclear gas kinematics probed by water vapor maser emission
has shown that two nearby active galaxies, NGC 1068 and NGC 4258,
have a supermassive object in their nuclei and
their masses are nearly comparable; a few 10$^7$
solar masses.
Nevertheless, the activity of the central engine of NGC 1068
is more powerful by two orders of magnitude than that of NGC 4258.
Since it is generally considered that
the huge luminosities of active galactic nuclei
are attributed to the mass accretion onto a supermassive
black hole,
the above observational results suggest that the accretion rate
in NGC 1068 is much higher than that in NGC 4258.
Comparing the kinematical properties
of the accreting molecular clouds between NGC 1068 and NGC 4258,
we find possible evidence for dynamical gas accretion
in NGC 1068, which may be responsible for the more
powerful central engine in this galaxy.
\end{abstract}



\section{Introduction}

NGC 1068 is the archetypical nearby 
active galaxy (cf.\ Antonucci, Miller 1985)
while NGC 4258 belongs to a class of
low-ionization nuclear emission-line regions (LINER; Heckman 1980).
The recent very-long-baseline interferometry (VLBI)
 measurements of H$_2$O maser emission of these two
galaxies  have shown that high-density molecular
gas clouds are orbiting around their central, massive object
(Miyoshi et al.\ 1995; Greenhill et al.\ 1995, 1996;
Gallimore et al.\ 1996a).
Although both the galaxies have a hidden
active galactic nucleus (AGN)
(Antonucci, Miller 1985; Wilkes et al.\ 1995),
their observational properties are quite different.
In table 1, we compare the observational
properties of the central engines and the molecular gas tori
in  NGC 1068 and NGC 4258.
The comparisons of hard X-ray, radio continuum,
and H$\beta$ luminosities
between them show that 
NGC 1068 is brighter intrinsically by two orders of magnitude
than NGC 4258.
In spite of this significant  difference in luminosity of
the central engines,
the estimated nuclear mass of
NGC 1068\footnote{The estimate 
of the nuclear mass
of NGC 1068: The VLBI measurement of H$_2$O maser 
emission by Greenhill et al.\ (1996)
gives the observed rotational velocity,
$v_{\rm rot} \simeq 300$ km s$^{-1}$,
 at the inner radius of the sub-Keplerian ring,
$r_{\rm in} = 0.56$ pc. 
Adopting an angle between the rotational axis of the torus
and the line of sight, $\phi \simeq 40^\circ$ (Young et al.\ 1996),
we obtain the nuclear mass, $M_{\rm nuc} \simeq  2.8
\times 10^7 M_\odot$
at distance 22 Mpc.} 
($M_{\rm nuc} \simeq 2.8 \times 10^7 M_\odot$)
is almost comparable to 
that of NGC 4258 ($M_{\rm nuc} \simeq 3.6 \times 10^7 M_\odot$).
It is usually considered that AGNs
are powered by an
accreting, supermassive black hole (Lynden-Bell 1969; Rees 1984).
In this model, the bolometric luminosity of central engines
depends solely on the mass accretion rate and thus there is
no dependency on the central black hole mass.
Therefore, the above observations suggest
that the mass accretion rate in NGC 1068
is much higher than that in NGC 4258. 
In order to understand what happens in both the nuclei,
we compare the dynamical properties of the masing
molecular tori between the
two galaxies and show possible evidence for dynamical
accretion of molecular gas
in NGC 1068.


\section{Discussion}

In order to consider a possible mechanism
which explains why NGC 1068 has a more powerful
AGN than NGC 4258, we compare the observational
properties of the H$_2$O masing clouds
between NGC 1068 and NGC 4258.
In figure 1, we show the schematic illustration of
the observed H$_2$O maser
emission of NGC 1068 (Greenhill et al.\ 1996) and NGC 4258 
(Miyoshi et al.\ 1995; Greenhill et al.\ 1995).
First, we summarize the observational properties of the molecular 
torus in NGC 4258.
The H$_2$O maser emission consists of the two components
(Miyoshi et al.\ 1995; Greenhill et al.\ 1995): 1) One is the main
component whose observed velocity corresponds to the systemic one.
This emission arises near the inner edge
of the torus along the line of sight.
This component is due to background amplified-beamed emission.
And, 2) the other is the high-velocity components which come from the
two tangential sections of the molecular torus. Since there is no
background continuum source for these components,
it is considered that they are due
to self amplification along the long gain
paths\footnote{X-ray irradiation
has been considered to be important for masing 
of the 22 GHz transition of water vapor
(Neufeld et al.\ 1994; 
Maloney et al.\ 1996; Wallin, Watson 1997).
In terms of this mechanism, typical path lengths enough to
cause the observed maser
emission would be of order 0.001 pc (Wallin, Watson 1997), 
being shorter than that stated in 
Miyoshi et al.\ (1995).
If each cloud has an diameter of 0.001 pc, the maser emission
could occur in a single cloud. However, if this is the case,
we could observe
maser emission form the entire ring, being inconsistent with
the observations.
In any case, in order to explain the observed high velocity components,
the argument on path length given in
Miyoshi et al.\ (1995) and Greenhill et al.\ (1995)
must be appreciated.}.
The typical 
line width of each maser emission line, $\sim 1$ km s$^{-1}$,
gives a path length of $\sim$  0.01 pc
(Miyoshi et al.\ 1995).
In summary, the masing emission of NGC 4258 
comes from these two portions of the
molecular torus. 
It should be mentioned that the masing emission is {\it not} observed from 
the shaded area shown in figure 1 because of either
the short path length
or no background continuum source
(Miyoshi et al.\ 1995; Greenhill et al.\ 1995).


Next we discuss the H$_2$O maser emission of NGC1068.
We should mention that the kinematical property of the maser emission
of NGC 1068 is significantly different from that of NGC 4258.
In NGC 4258, the high-velocity components  are seen
only in the portions of 
Keplerian ring where the path length is long enough to
achieve self-amplification
(Miyoshi et al.\ 1995; Greenhill et al.\ 1995).
On the other hand, in the case of NGC 1068,
there appear
many masing clouds in the ``{\it unseen}'' area
in which self amplification cannot be achieved;
e.g., the component No.\ 4
and the majority of component No.\ 3
identified by Greenhill et al.\ (1996).
In order to elucidate difference in kinematical property of the masing clouds 
between NGC 1068 and NGC 4258,
we estimate quantitatively 
the unseen area of masing tori in both galaxies.
Since a long path length ($\simeq $ 0.01 pc; Miyoshi et al.\ 1995)
is necessary to cause the self-amplified  masing
in a molecular torus with the (sub-) Keplerian rotation,
most parts of the molecular ring become to be the unseen area
 (Miyoshi et al.\ 1995; Greenhill et al.\ 1995).
We consider a molecular torus with a radial velocity distribution of
$v_{\rm c}(r) = v_{\rm in} ~ (r/r_{\rm in})^a$
where $v_{\rm c}(r)$ is the circular velocity
at radius $r$ and $v_{\rm in}$ is the circular velocity
at the inner edge of the torus ($r_{\rm in}$) [see figure 1].
The power index, $a$, is $-0.5$ for the Keplerian torus
(i.e., for NGC 4258)
while $-0.31$ for NGC 1068 (Greenhill et al.\ 1996).
The velocity gradient along the line of sight is
given by the derivative,

\begin{equation}
{d v_{\rm l}(r) \over d y}  =  {1 \over 2}~ (a - 1) ~
\omega_{\rm c}(r) ~ \sin 2\theta 
\end{equation}

\noindent where $v_{\rm l}(r) = v_{\rm c}(r) \sin \theta$ is the
line of sight velocity at radius $r$ and
$\omega_{\rm c}(r) =  v_{\rm c}(r) / r$. 
The definition of $\theta$ is shown in figure 1.
In the unseen area,
the velocity change ($\Delta v$) within a path length ($l$)
should be larger than
1 km s$^{-1}$ (Miyoshi et al.\ 1995);
i.e., $\vert dv_{\rm l}(r) / dy \vert  >
\Delta v / l$. This condition gives a constraint
on the unseen angle, $\theta$,

\begin{equation}
\sin 2\theta  ~ > ~ {2 \Delta v/l \over \vert a - 1
\vert ~ \omega_{\rm c} (r)}. 
\end{equation}

\noindent In the lower panel of figure 1,
we show the unseen area for both NGC 4258 and NGC 1068.
Here we adopt $(v_{\rm in}, r_{\rm in})$ =
(1080 km s$^{-1}$, 0.13 pc) for NGC 4258 and
$(v_{\rm in}, r_{\rm in})$ = (300 km s$^{-1}$, 0.56 pc) for NGC 1068.
In the case of NGC 4258, it is shown that the high velocity component
can arise from the tangential sections of the torus,
being consistent with
the observations (Miyoshi et al.\ 1995; Greenhill et al.\ 1995).
As for the main component,
if we assume the point-like continuum radiation source,
we need an opening angle
of 7$^\circ$ (Miyoshi et al.\ 1995).
 Since, however, this angle is larger than the critical unseen
angle (0$^\circ$\hspace{-4.5pt}.\hspace{.5pt}5),
we consider that the continuum
source may have  a spatial extent of
$\sim$ 0.01 pc (cf.\ Haschick, Baan 1994).
Although the unseen area estimated for the case of NGC 1068 is
basically similar to that of NGC 4258, the observed appearance of 
the masing clouds in NGC 1068 is significantly different from that 
of NGC 4258.
Therefore, we must consider some other different masing mechanisms for the 
masing clouds of NGC 1068 which appear in the unseen area.
 
Greenhill et al.\ (1996) proposed that the maser emission
comes from the limb of the torus rather than the midplane.
Though they argued that the orbital motion is parallel to
the line of sight along the limb and produces a 
substantial amplification.
However, since the limb may co-rotate with the midplane
(i.e., the sub-Keplerian rotation), it is unlikely
that only the gas clouds
on the limb can gain the long path.
Therefore, the masing regions
cannot be attributed to the sub-Keplerian rotating molecular  torus.
It is thus suggested that they come from dense molecular clouds inside
the torus\footnote{Begelman, Bland-Hawthorn (1997) reported that
the masing clouds in the nuclear region 
of NGC 1068  are distributed in the warped disk.}. 
In order for such clouds to gain the long path length,
the velocity structure should obey a rigid-body rotation.
There are two alternatives for such a cloud system. One is a gaseous
bar (or an oval ring) with a figure rotation.
Since the sub-Keplerian torus suggests the presence of
gravitational disturbance to this torus,
it is likely that there is the inner non-axisymmetric potential
like a bar structure.
The other possibility is that clumpy molecular clouds
rotate like a rigid body as a whole.
A typical size of  each cloud
 would be as large as $\sim$ 0.01 pc to gain
the long path length enough to cause the masing.
However, each cloud itself may 
rotate because of the conservation of angular momentum.
Thus if this is the case, each cloud would rotate as a rigid body 
rotator again.

Here a key question arise as how to make such
a gaseous bar or a clumpy cloud system
inside the torus.
The gradual accumulation of gas clouds toward the central region of the
galaxy causes over density in the gas cloud system.
Once the gas mass exceeds about one tenth of the
dynamical mass within the concerned region, the gas may experience
the gravitational instability because of its self gravity, 
leading to the formation of either
a bar (Shlosman et al.\ 1989;
Shlosman et al.\ 1990; 
Wada, Habe 1995)  or clumpy gas clouds 
(Shlosman, Noguchi 1993; Heller, Shlosman 1994).
If the gas mass within $r \simeq$ 0.56 pc would 
exceed $\sim 3 \times 10^6 M_\odot (\sim 0.1 \times M_{\rm nuc})$,
the gravitational instability may occur in the nuclear gas of NGC 1068.
Although, however,  NGC 1068
has abundant dense gas in the nuclear region (Tacconi et al.\ 1994),
the gas mass associated with the dusty torus is estimated to be
$\sim 3 \times 10^4 M_\odot$ at most (Pier, Krolik 1992b, 1993).
It is therefore unlikely that the rigid-body component
comes from the gravitational instability of the torus itself.
Taking the more powerful central engine of NGC 1068 into account,
we may consider that the inner radius of the torus in NGC 1068
would be once much smaller. If this is the case, we expect that
the inner parts of the molecular torus may be broken by the intense
radiation field (cf.\ Pier, Krolik 1992a;
see also Neufeld, Maloney 1995). 
Although we cannot specify the actual geometry of the 
rigid-body component,
we show our modest interpretation schematically in figure 2.


Finally we estimate a possible gas accretion rate from
the inner torus broken by the intense radiation in NGC 1068. 
Since it is considered that the gas clouds broken by
the radiation may accrete in a characteristic time
which is roughly comparable to the free fall time there,
$t_{\rm ff} = \sqrt{r^3/(G M_{\rm nuc})} \sim 10^3$ years
for $r=0.5$ pc.
Since, however, we do not take the angular momentum loss into
account in this estimate, this timescale should be considered
as a lower limit.
The gas mass broken by the radiation may be the same order
as that of the present torus, $M_{\rm cloud} \sim 10^4 M_\odot$,
giving a nominal {\it dynamical} accretion rate as
${\dot M}_{\rm acc} ({\rm Dyn}) \sim M_{\rm cloud} / t_{\rm ff}
\leq 10 M_\odot$ y$^{-1}$.
Though it is hard to estimate the effect of angular momentum loss accurately
because of ambiguity in the physical and dynamical condition of the
nuclear gas, we may estimate the probable accretion timescale 
from the actual size of the narrow line region (NLR) in NGC 1068.
Given the size of the NLR, $r_{\rm NLR} \simeq$ 900 pc 
at the distance $D$ = 22 Mpc (Schmitt, Kinney 1996),
the timescale (i.e., the lifetime of the
central engine)  is estimated to be $t_{\rm acc} \sim r_{\rm NLR}/(\beta c)
\sim 3 \times 10^4 / (\beta/0.1)$ years where  $\beta c$ is the net velocity
of the ionizing radiation ($c$ is the light velocity in the vacuum). 
Here we assume $\beta \sim 0.1$
because it is unlikely that the ionizing radiation is highly
relativistic (cf.\ Gallimore et al.\ 1996b).
Since this timescale is more reliable than that estimated above,
we obtain the most probable dynamical accretion rate,
${\dot M}_{\rm acc} ({\rm Dyn}) \sim 0.3 (\beta / 0.1) M_\odot$ y$^{-1}$.
Next, we estimate another accretion rate based on the bolometric 
luminosity.
Pier et al.\ (1994) gave the bolometric luminosity of
the central engine in NGC 1068,
$L_{\rm bol} \simeq
8.5 \times 10^{44}~ (f_{\rm refl}/0.01)^{-1} (D/22 {\rm Mpc})^2$
erg s$^{-1}$ where
$f_{\rm refl}$ is the fraction of nuclear flux reflected into our line of sight
and $D$ is the distance to NGC 1068. Adopting  the fiducial values
in Pier et al.\ (1994), we obtain the gas accretion rate
based on the luminosity,
${\dot M}_{\rm acc} ({\rm Lum}) 
= L_{\rm bol} / (\eta_{\rm acc} c^2)  \simeq 0.02 ~
(L_{\rm bol}/ 10^{44} ~ {\rm erg~ s}^{-1})
(\eta_{\rm acc}/ 0.1)^{-1}
\simeq 0.17 ~ M_\odot$ y$^{-1}$
where $\eta_{\rm acc}$ is the conversion efficiency from
the gravitational energy to the radiation.
Since this value is almost consistent with the dynamical one,
we consider that the dynamical accretion may be responsible for
the activity in NGC 1068.

In conclusion, the striking difference of the molecular gas tori
between NGC 1068 and NGC 4258 is the 
presence/absence of a rigid-body rotating component which may be formed
through the destruction of the innermost torus
by the strong radiation from the central engine.
This component may provide direct evidence for dynamical accretion 
of the nuclear gas in NGC 1068.
Since this component is not observed in the less-luminous AGN
of NGC 4258,
we suggest that this component 
is  responsible for the more powerful AGN
in NGC 1068.
\par
\vspace{1pc} \par
We would like to thank Lincoln Greenhill and Naomasa Nakai for
helpful discussion about the H$_2$O maser in NGC 1068 and  NGC 4258.
We also thank Kazuo Makishima for useful comments of the X-ray property
of NGC 4258 and the referee, Jun Fukue, for kind comments
which improved 
this paper.
This work was supported by
the Ministry of Education, Science, Sports, and Culture
(No.\ 07044054).
TM was supported by the Grant-in-Aid for JSPS Fellows
by the Ministry of Education, Science, Sports, and Culture.

\clearpage


\clearpage
\begin{table*}[t]
\begin{center}
Table~1.\hspace{4pt}Comparison of NGC 1068 and NGC 4258.\\
\vspace{6pt}
\end{center}
\begin{tabular*}{\textwidth}{@{\hspace{\tabcolsep}
\extracolsep{\fill}}p{15pc}ccc}
\hline\hline
    & Unit & NGC 1068 & NGC 4258 %
\\\hline
Distance \dotfill & Mpc & 22 (1) & 6.4 (2) \\
Central mass ($M_{\rm nuc}$) \dotfill & $M_\odot$ &
          $2.8 \times 10^7$ (3) & $3.6 \times 10^7$ (2) \\
H$\beta$ luminosity$^\dagger$ ($L_{\rm H\beta}$) \dotfill &
          erg s$^{-1}$ & $2.0 \times 10^{42}$ (4) &
          $1.4 \times 10^{40}$ (5) \\
X-ray luminosity$^{\dagger\dagger}$ ($L_{\rm X}$) \dotfill &
          erg s$^{-1}$ &  10$^{43 {\rm -} 44}$ (6) & $3.3 \times 10^{40}$ (7) \\
20 cm radio power ($P_{\rm 20 cm}$) \dotfill & W Hz$^{-1}$ & 
         2.2$\times 10^{23}$ (8) & $4.0 \times 10^{21}$ (9) \\
H$_2$O maser luminosity ($L_{\rm H_2O}$) \dotfill &
         $L_\odot$ &  145 (10) & 125 (10) %
\\\hline
\multicolumn{4}{c}{Molecular torus} %
\\\hline
Inner radius ($r_{\rm in}$) \dotfill & pc & 0.56 (11) & 0.13 (2) \\
Outer radius ($r_{\rm out}$) \dotfill & pc & 1.0 (11) & 0.25 (2) \\
Inner rotation velocity ($v_{\rm in}$) \dotfill & km s$^{-1}$ &
              300 (11) & 1080 (2) \\
Inclination \dotfill & degree & $\sim 40 \pm 5$ (12) & $83 \pm 4$ (2) \\
Position angle \dotfill & degree & $\sim 45$ (11) & $86 \pm 2$ (2) %
\\ \hline
\end{tabular*}
\vspace{6pt}\par\noindent
$^\dagger$ The intrinsic luminosity of the hidden,
broad H$\beta$ emission;
$L_{\rm H\beta}$ = $L_{\rm H\beta}^{\rm p}
(\Delta\Omega/4\pi)^{-1}$ $\tau^{-1} P^{-1}$
where $L_{\rm H\beta}^{\rm p}$ is the polarized
H$\beta$ luminosity, $\Delta\Omega$
is the covering factor in steradian, $\tau$ is the
scattering optical depth, and $P$
is the fractional polarization (Miller, Goodrich 1990).
We adopt ($\tau \Delta\Omega/4\pi$, $P$) =
(0.034,  0.16) for NGC 1068 and (0.034, 0.094) for NGC 4258.
Here we assume that
NGC 4258 has the same value of $\tau \Delta\Omega/4\pi$ as
that of NGC 1068.
The intrinsic H$\beta$ luminosity is considered
to be proportional to the
bolometric luminosity.
\vspace{6pt}
\par\noindent
$^{\dagger\dagger}$ The reddening-corrected 
hard X-ray luminosity (2 - 10 keV).
\vspace{6pt}
\par\noindent
Refs.: 1) This value is usually adopted, corresponding to the recession
velocity 1,100 km s$^{-1}$ with a Hubble constant
$H_0$ = 50 km s$^{-1}$ Mpc$^{-1}$.
2) Miyoshi et al.\ (1995). 3) This paper (see the footnote).
4) Miller and Goodrich (1990). 5) Wilkes et al.\ (1995). 
6) Koyama et al.\ (1989). 7) Makishima et al.\ (1994). 
8) Ulvestad and Wilson (1984), 9) Hummel et al.\ (1989).
10) Nakai et al.\ (1995).
11) Greenhill et al.\ (1996).
And, 12) Young et al.\ (1996). Note that this value is obtained
not for the molecular torus but for the dusty one.
\end{table*}

\clearpage
\centerline{Figure Captions}
\bigskip
%
\noindent
{\bf Fig.\ 1: }
{The unseen area of masing tori
for NGC 4258 and NGC 1068.}
%

\bigskip
\noindent
{\bf Fig.\ 2: }
{Schematic comparison of the molecular tori
of NGC 1068 (Greenhill et al.\ 1996) and NGC 4258 
(Miyoshi et al.\ 1995; Greenhill et al.\ 1995) probed by
the H$_2$O maser emission.
Although the rotation of the actual molecular torus of NGC 4258 is
clockwise, we adopt the counterclockwise rotation for both types
nominally.}
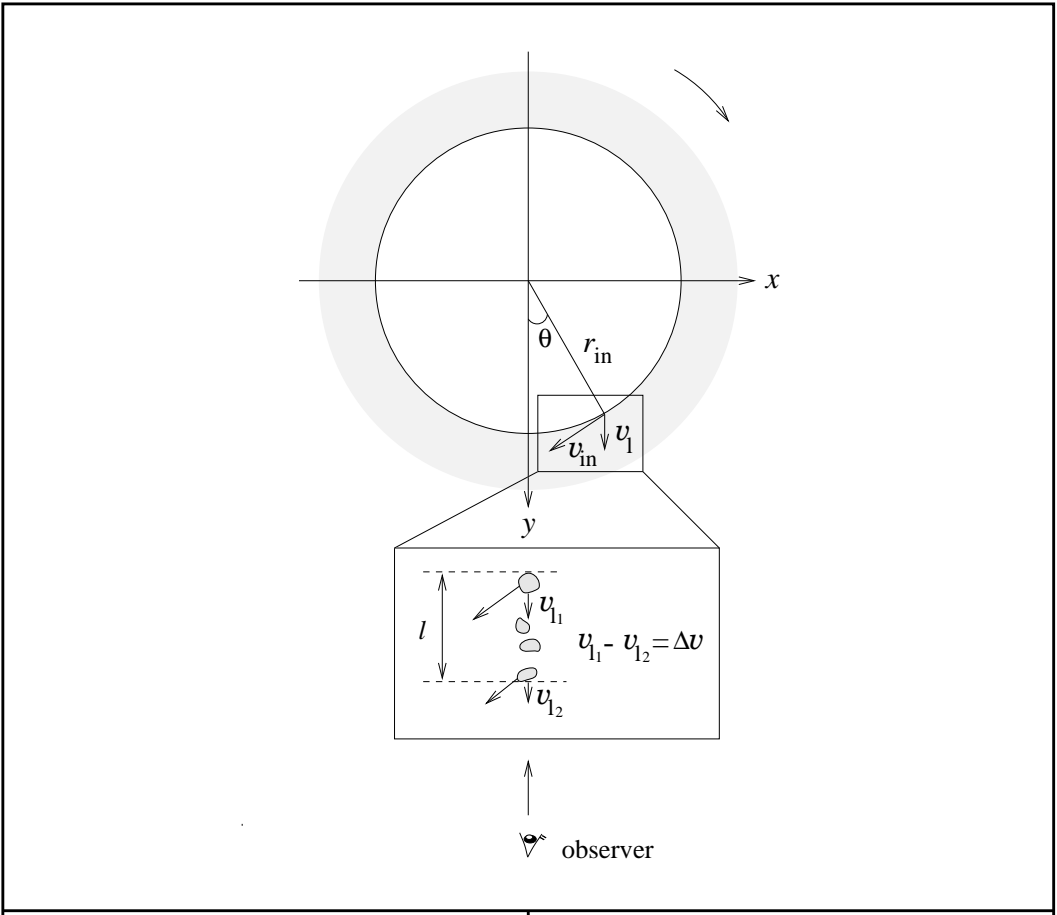
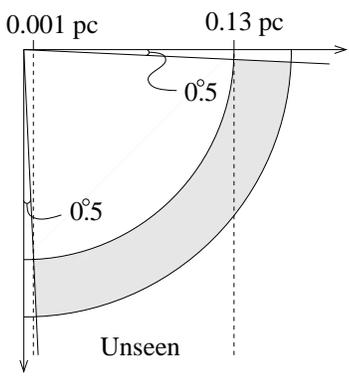
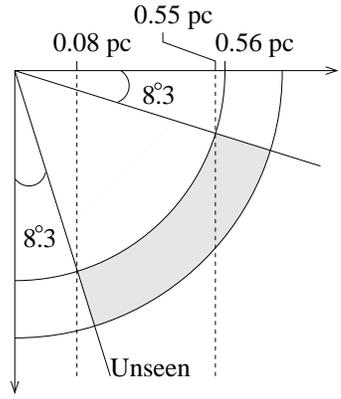
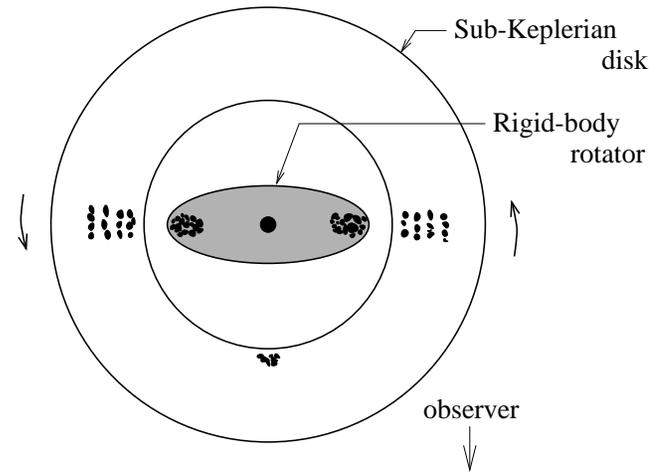
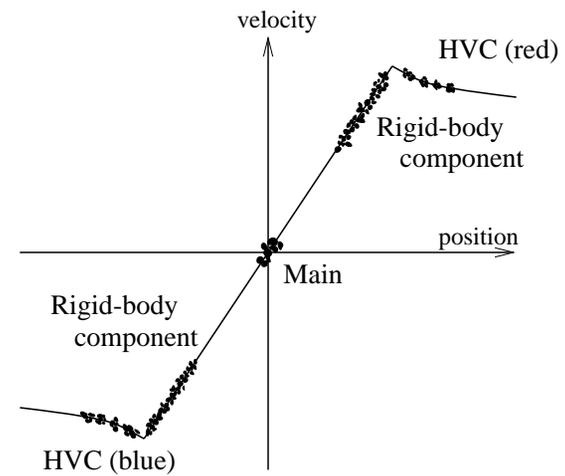
\end{document}